\documentclass{ws-procs9x6}

\begin{document}

\title{Kinks in Time and Their Relation to Confinement
\footnote{\uppercase{W}ork partially supported by
\uppercase{NSC}, \uppercase{NCTS} and the
\uppercase{C}os\uppercase{PA} project.}
}


\author{John E.~Wang}

\address{Jefferson Physical Laboratory, Harvard University, Cambridge, MA
02138, USA \\
Department of Physics, National Taiwan University,
Taipei 106, Taiwan\\
E-mail: hllywd2@feynman.harvard.edu, hllywd2@phys.ntu.edu.tw}




\maketitle

\abstracts{The time dependent formation of an electric flux tube
(fundamental string) is reviewed.  The main tool used for analysis
is the Spacelike brane, which is a kink in time of the rolling
tachyon. Both the S-brane and rolling tachyon are attempts to
extend the D-brane concept to time dependent backgrounds. While
S-branes are similar to Euclidean counterparts of the more
familiar timelike D-branes, S-branes can smoothly change their
worldvolume signature from spacelike to timelike which we
interpret as the formation of a topological defect.}






Here we review the results of Ref.~1
(see also references therein), in particular the derivation of
actions and construction of solutions of Spacelike
branes\cite{GS} which are useful in understanding time dependent
systems such as the rolling tachyon\cite{Sen-RT}.

Recent proposals to understand the dynamics of our universe's
cosmic acceleration include searching for naturally unstable and
hence time dependent backgrounds corresponding to extensions of
the fruitful D-brane concept.  The motivation is that the study
of BPS branes has been one of the notable recent successes,
providing a driving force in string theory and proving immensely
useful in the study of supersymmetric and static systems.  To
develop our tools and intuition in time dependent backgrounds we
are led to consider non-supersymmetric branes.

One such unstable background involves brane and anti-brane pairs,
and their closely related non-BPS branes.  Put close together, a
brane and anti-brane pair will naturally interact and can
annihilate, and this process can be understood as tachyon
condensation. An unstable $p$-brane carries a scalar field $T$
called the tachyon field governed by a tachyon potential which is
approximately of the form $V(T)\approx e^{-T^2}$. The tachyon
field parametrizes the instability in the system. When the
tachyon value is near zero, the brane system is unstable and will
be driven towards the bottom of the tachyon potential at large
tachyon values.

We begin our study of the dynamical evolution of the tachyon
condensation process by examining the rolling tachyon solution,
which is a time dependent and spatially homogenous tachyon
solution
\begin{equation}
T=T(t=time) \ .
\end{equation}
The tachyon values can start off near zero and then accelerate
towards large tachyon values in the same way a ball rolls down a
hill.  At the top of its trajectory the tachyon field has value
$T=0$.  This $p$ dimensional Euclidean hypersurface corresponding
to the moment in time when the tachyon scalar field is everywhere
zero we call a Spacelike $(p-1)$-brane.  As it stands this
Euclidean hypersurface does not carry much dynamics as it appears
and disappears in an instant leaving no apparent trace of its
existence.  Despite its spacelike trajectory, it clearly does not
violate causality nor does it transport energy superluminally.




To extend the S-brane lifetime we deform the S-brane worldvolume
by introducing fluctuations into the rolling tachyon profile. The
deformed S-brane now lives for a finite period of time and is not
homogenous and as finely tuned.  If we make an infinite
deformation of the S-brane, it will live for an infinite length
of time as shown in Figure~1.

\begin{figure}[ht]
\centerline{\epsfxsize=3.5in\epsfbox{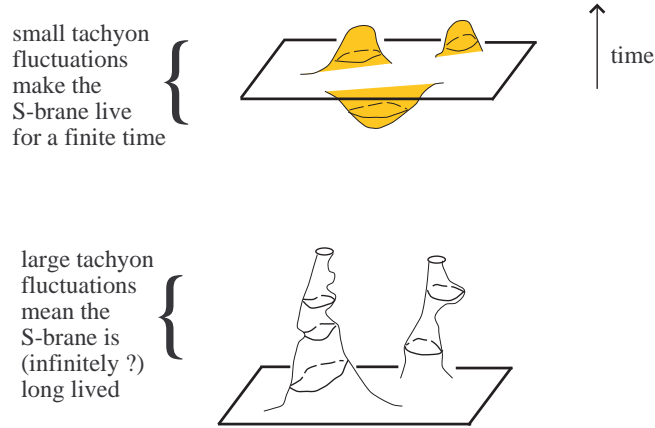}}
\caption{A flat S-brane can be deformed and live for more than a
moment in time.  Time increases as we go up the page and a
horizontal plane is a moment in time. }
\end{figure}

The key point is that large fluctuations do naturally occur and
they correspond to the time dependent formation of topological
defects.  S-branes can be thought of as initial conditions for
the formation of defects.

To clarify why S-branes play this role in defect formation, we
recall that for static tachyon kink configurations the zeros of
the tachyon field
\begin{equation}
T=0
\end{equation}
are interpreted as the location of lower dimensional solitons or
topological defects. In the case of string theory these solitons
include branes and anti-branes.  This is why in analogy, we
previously identified the $T=0$ region of a rolling tachyon or
kink in time to be the location of the Spacelike brane.

The time evolution of defect formation can be followed and traces
out a spacetime trajectory which is the S-brane worldvolume. In
other words because both S-branes and normal defects are
specified by the condition $T=0$, slow moving and long lived
Spacelike branes should be able to represent ordinary and stable
timelike topological defects! (See Figure~2.)

\begin{figure}[ht]
\centerline{\epsfxsize=3.9in\epsfbox{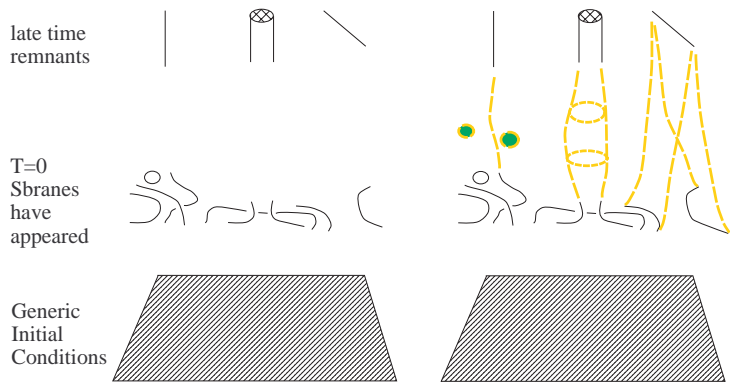}}
\caption{S-branes worldvolumes (dashed lines) allow us to follow
the time evolution of initial conditions during the formation of
topological defects.}
\end{figure}

To quantify this statement we have derived an effective action for
Spacelike branes by performing a zero mode fluctuation analysis on
non-BPS branes.  The S-brane action is

\begin{equation}
S  =  \int_{world\ volume} \sqrt{det(\delta_{ij}-\partial_i t
\partial_j t + F_{ij})} \ .
\end{equation}

This action passes the following tests (Ref.~1 discusses further
checks)

1)there is a scalar field, $t$, with wrong sign kinetic energy
corresponding to fluctuations in time

2)gauge fields with field strength $F_{ij}$ exist on the
worldvolume just like for D-branes

3)the action is real for flat Euclidean hypersurfaces of equal
time

4)yet infinitely long lived solutions which smoothly change from
Euclidean to Lorentzian signature are also possible. This property
is based on the fact that the S-brane action is by definition
defined to be spacelike relative to the open string metric whose
light cone always lies within the closed string light
cone\cite{lightcone}, $G_{open}^{\mu\nu} \subseteq
G_{closed}^{\mu\nu}$. There are therefore long lived solutions
which are consistently both spacelike relative to the open string
metric and also timelike relative to the closed string metric as
illustrated in Figure~3.

\begin{figure}[ht]
\centerline{\epsfxsize=2in\epsfbox{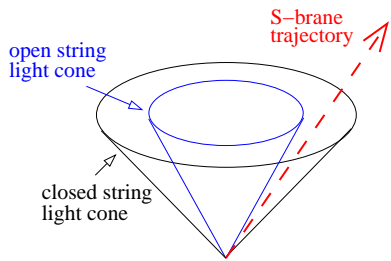}}
\caption{S-branes can simultaneously be spacelike relative to the
open string metric and timelike relative to the closed string
metric.  The narrower the open string light cone becomes, the
slower the S-brane is allowed to travel.}
\end{figure}

5)solutions exist describing the formation of an electric flux
tube

Although static descriptions of fundamental strings have been
discussed\cite{volcanos}, this S-brane solution describes the
time dependent formation of a confined electric flux tube (see
also Ref.~6). In gauge theories the bundling of gauge fields into
small regions is the process of confinement, so this solution
indicates that confinement can be understood as a
non-perturbative but classical dynamical process.  As an example,
the electric S3-brane is cylindrical in shape, $\mathbb{R}^1\times
S^2$, where $r$ is the radius of the cylinder, $\chi$ goes
lengthwise along the cylinder and $t$ is time.  In particular,
this electric S3-brane solution is
\begin{equation}
r= \frac{c}{t}, \hspace{.3in} E = 1
\end{equation}
where the electric field, $F_{0\chi}=E$, is constant and the
critical value.  The radius of this cylinder shrinks to zero at
late times while the electric field along the cylinder is
constant, so this solution describes an electric flux tube which
is confining into a string-like object. Using the Dirac
quantization condition we found that the energy is that of a
fundamental string
\begin{equation}
H = \frac{1}{2\pi \alpha^\prime} \int d\chi \left( \int d^px D
\right)= \frac{1}{2\pi \alpha^\prime} \int d\chi \ .
\end{equation}
Further, the S-brane coupling to RR form fields, $A$, goes to zero
\begin{equation}
\int_{world \ volume} A \times (t^{-1})
\stackrel{{t\rightarrow \infty}}{\longmapsto} 0 \ . \\
\end{equation}

\noindent so the D-brane charge vanishes.  The time dependent
factor of $t^{-1}$ is due to the fact that the worldvolume of the
S-brane shrinks to zero.


Spacelike branes are extensions of the D-brane concept to time
dependent backgrounds and are useful in understanding the
formation of defects. From the S-brane perspective, confinement
can appear as a classical self organizing process and might lead
to a hint as to the physics at the Hagedorn temperature.
Spacelike branes should play a similar role in time dependent
backgrounds of ordinary field theories.

\section*{Acknowledgments}
It is a pleasure to thank Koji Hashimoto, Pei-Ming Ho and Satoshi
Nagaoka for wonderful collaboration.  I also wish to thank the
organizers for their kind invitation to the stimulating
Confinement 2003 conference.


\end{document}